\documentclass{article}
\usepackage{spconf,amsmath,graphicx,tabularx,hyperref}

\usepackage{enumitem,booktabs}
\setlist{nosep, leftmargin=14pt}

\usepackage{mwe} 
\usepackage{multirow}

\usepackage{etoolbox}
\usepackage{booktabs}

\title{Lightweight Framework for Automated Kidney Stone Detection\\ using coronal CT images}
\name{Fangyijie Wang$^{\dagger}$ \qquad Gu\'enol\'e Silvestre$^{\star}$ \qquad Kathleen M. Curran$^{\dagger}$}

 \address{$^{\dagger}$School of Medicine, University College Dublin, Dublin, Ireland \\
    $^{\dagger}$fangyijie.wang@ucdconnect.ie \\
     $^{\star}$School of Computer Science, University College Dublin, Dublin, Ireland}
\begin{document}
%
\maketitle
\begin{abstract}
Kidney stone disease results in millions of annual visits to emergency departments in the United States. Computed tomography (CT) scans serve as the standard imaging modality for efficient detection of kidney stones. Various approaches utilizing convolutional neural networks (CNNs) have been proposed to implement automatic diagnosis of kidney stones. However, there is a growing interest in employing fast and efficient CNNs on edge devices in clinical practice. 
In this paper, we propose a lightweight fusion framework for kidney detection and kidney stone diagnosis on coronal CT images. In our design, we aim to minimize the computational costs of training and inference while implementing an automated approach. The experimental results indicate that our framework can achieve competitive outcomes using only 8\% of the original training data. These results include an F1 score of 96\% and a False Negative (FN) error rate of 4\%. Additionally, the average detection time per CT image on a CPU is 0.62 seconds.

Reproducibility: 
Framework implementation and models available on GitHub. 
\end{abstract}
\section{Introduction}
\label{sec:intro}

Kidney stone disease is a highly prevalent condition worldwide. It has the potential to result in kidney failure, reduce workforce productivity due to the intense pain it causes, and negatively impact the quality of life by obstructing the urinary system~\cite{Penniston_2012}. Recent research found that the prevalence of kidney stones in the United States has significantly increased, rising from 3.2\% in 1980 to 10.1\% in 2016~\cite{Chewcharat_2021}. Therefore, research on kidney stone disease has the potential to enhance patients' quality of life and lower the risk of kidney failure.

CT has become the standard imaging technique for the diagnosis of kidney stones~\cite{Turk_2015}. In recent years, Deep Learning (DL) has showcased its prowess in extracting textural patterns and morphological features from medical images for analysis. Consequently, it has emerged as a leading approach, achieving state-of-the-art performance in various medical applications~\cite{Shen_2017}. Recent works~\cite{Hesamian_2019, talo_2019, ozturk_2020, Kijowski_2020} demonstrated successful applications across various fields, assisting clinicians in disease diagnosis with DL methods. Simultaneously, there is a growing expectation among doctors and patients for medical images to be easily accessible on edge devices, enabling consultative viewing~\cite{hirschorn_2014}.

In this paper, we present a lightweight deep learning (DL) framework for efficient kidney stone detection on coronal CT images. The primary objective is to optimize time consumption and minimize the error rate associated with kidney stone detection, even with a limited number of samples. Our contributions are as follows: (i) A lightweight fusion framework of YOLO~\cite{Redmon_yolo_2016, Jocher_2023} and the convolutional neural network (CNN) classifier is proposed for the analysis of CT images; (ii) Enhancement of kidney stone classification by extracting the region of interest (ROI); (iii) A low-cost training approach for the detection of kidney stones in CT scans with a public release of model implementation available on GitHub. 

\section{Related Work}
\label{sec:rel_work}

The first work~\cite{Yildirim_2021} using coronal CT images~\cite{Yildirim_2021} for automatic kidney stone detection proposed a DL model with the cross-residual network architecture~\cite{Jou_2016}. The model yielded a 96.8\% accuracy rate and a 95.8\% sensitivity using 146 test cases and 346 images. The results showed that the DL model successfully identified the areas of interest in the decision-making process. Clinically, the regions identified by the DL model were consistent with the evaluations made by medical experts for the majority of the images.

The recent work~\cite{Islam_2022} demonstrates the effectiveness of the transformer model in the field of diagnosis using a large dataset of CT images. Specifically, the Swin transformer~\cite{liu_2021} model outperformed other topologies such as VGGNet~\cite{Simonyan_2015}, ResNet~\cite{he_2016}, and Inception~\cite{Szegedy_2015} in terms of accuracy, achieving an accuracy rate of 99.3\%. This improvement is also observed in other metrics such as F1, precision, or recall. It is important to note that these impressive results were obtained after training on a substantial CT dataset consisting of 12,446 whole axial and coronal CT images. Another recent study~\cite{baygin_2022}, presenting novel findings, introduced a fused model that combines DL and feature engineering, referred to as ExDark19, to achieve superior classification performance. 

The latest study~\cite{Bhandari_2023} reported an average accuracy of 99.52\% ($\pm$ 0.84) on kidney stone detection using a specially designed lightweight CNN model. This work suggests that small footprint models could be deployed on Internet of Medical Things (IoMT)-enabled devices without sacrificing performance. Unlike this previous study, our research focuses on the utilization of small-sample training methods by employing lightweight DL models specifically designed for diagnostic purposes on coronal CT images. The previous study~\cite{Metser_2009} demonstrated coronal images' ability to aid physicians in detecting more positive cases than axial images.

\section{Materials and Methods}
\label{sec:method}
We propose to leverage the YOLO model~\cite{Redmon_yolo_2016, Jocher_2023} to extract the ROI from kidney CT images, and a lightweight CNN model to classify the kidney stone from ROI. Both models are reviewed in Section~\ref{ssec:YOLOv8n}.

\subsection{Dataset}
\label{ssec:Dataset}
The CT dataset~\cite{Yildirim_2021} used in this study is a public collection consisting of images acquired in the supine position using a single scanner Philips Healthcare, Ingenuity Elite (Netherlands).
The radiologists and urologists classified the presence of stones without performing any segmentation on the CT images. 
A total of 433 subjects were included in this dataset, with 278 being stone positive and 165 being normal. The study utilized CT images obtained from different sections of these patients, resulting in a collection of 790 images for patients diagnosed with kidney stones and 1009 images of normal subjects.


To prevent biased results, subject data used during the training and validation phases were excluded from the testing phase. The CT images were divided into 3 sets: the training set (1163 images), the validation set (290 images), and the test set (346 images). We randomly selected 100 images from the training set to train the YOLOv8n model~\cite{Jocher_2023} for kidney detection. After training and processing, the model generated two predictions of kidneys on each of these images, resulting in 200 
processed output. From these kidney images, another set of 100 was randomly selected to train a classifier model based on the MobileNet V2 architecture~\cite{Sandler_2018}. The remaining 100 kidney images were kept for testing. 

\textbf{Data Augmentation.} With a training set of 100 CT images used for kidney stone classification, we employed various data augmentation techniques during  the training phase. Consequently, our investigation aimed to assess the potential of these data augmentation techniques in improving the accuracy of CNN models. 

We analyzed the distribution of the target label in the training set prior to commencing the classification and observed that labels exhibited an imbalanced distribution, with 137 stone-positive images and 63 normal images. To address the imbalance issue, we exclusively applied data augmentation techniques to the stone-positive images within the training set. This increased the number of stone-positive images. There are three data augmentation techniques we employed in this study: (i) horizontally flipping each image, (ii) randomly rotating half of the kidney images in the range of $-25^\circ$ to $25^\circ$, and (iii) pixel normalization~\cite{Olga_2015} (RGB mean and standard deviation set to $[0.485, 0.456, 0.406]$ and  $[0.229, 0.224, 0.225]$ resp.\footnote{These values were obtained from pre-trained models on the ImageNet dataset, ensuring accuracy and reliability.}). The last technique was applied to all CT images within the training set and test set. As a result of these augmentation techniques, the size of the training set expanded to 293 instances.

\subsection{Proposed Fusion Framework}
\label{ssec:model_pipe}

\begin{figure}[htb]

\begin{minipage}[b]{1.0\linewidth}
  \centering
  \centerline{\includegraphics[width=8.5cm]{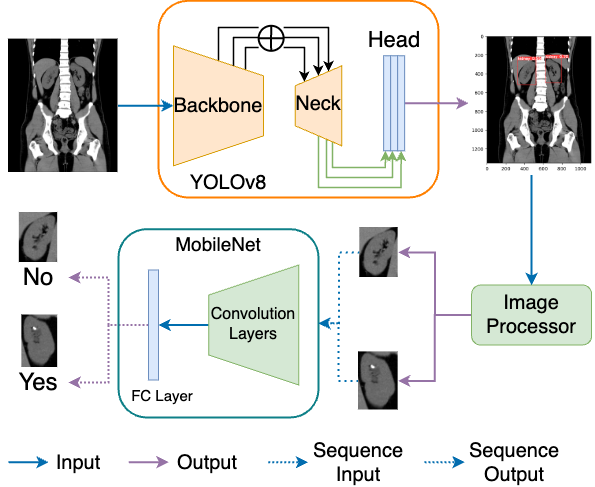}}
\end{minipage}
\caption{General overview of the proposed workflow and framework for kidney stone detection and classification.}
\label{fig:model_pipe}
\end{figure}

The structure of our model pipeline is depicted in Fig.~\ref{fig:model_pipe}, and it comprises three main components: the YOLOv8 model~\cite{Jocher_2023}, Image Processor, and MobileNet classifier~\cite{Sandler_2018}. In our pipeline, we employ the YOLOv8 Nano model, which is the smallest and fastest model in the YOLOv8 family. Table~\ref{yolo_family} illustrates the comparison of models within the YOLOv8 family. Despite YOLOv8 Nano exhibiting the lowest accuracy (validatian mAP: 50--95) on the COCO dataset~\cite{lin_2014}, we believe that it is well capable of detecting kidneys in CT images, as reported in previous research~\cite{Mahmud_2023}. The purpose of using the YOLO model is to segment the main region of interest (ROI) containing human kidneys in all CT scans. Our assumption is that kidney stones only occupy a small area in term of pixels in coronal CT scans. In other words, we believe that the detection and classification of kidney stones are challenging due to their small size in coronal CT scans and their densities being similar to that of bones. Therefore, by successfully segmenting most of the ROIs, it is likely that the segmented areas will also include the pixels corresponding to kidney stones. Subsequently, a CNN in the downstream can classify the kidney stones using only the ROIs as input images.

The image processor is a Python script 
to extract and crop the ROIs from predicted images using the red bounding boxes output of YOLOv8n. Consequently, a significant amount of irrelevant information for kidney stone classification is disregarded during the processing stage. The primary goal is to retain only the ROIs related to kidneys and kidney stones, excluding areas with similar densities as human bones.

\begin{table}[htbp]
    \caption{Evaluation details of YOLOv8 models trained on the COCO dataset. Model suffix 'n' stands for nano, 's' for small, 'm' for medium, 'l' for large, and 'x' for extra large.}
    \vspace*{0.5em}
    \renewcommand{\arraystretch}{1.3}
    \setlength\tabcolsep{1ex}
    \begin{tabularx}{\linewidth}{
        >{\centering\arraybackslash}p{0.12\linewidth}
        >{\centering\arraybackslash}p{0.1\linewidth}
        >{\centering\arraybackslash}p{0.12\linewidth}
        >{\centering\arraybackslash}p{0.1\linewidth}
        >{\centering\arraybackslash}p{0.1\linewidth}
        >{\centering\arraybackslash}p{0.08\linewidth}
        >{\centering\arraybackslash}p{0.1\linewidth}
        }
        \hline
        \textbf{YOLO Model} & \textbf{Size} (pixel) & \textbf{Val. mAP 50-95}$\downarrow$ & \textbf{Speed CPU} (ms)$\downarrow$& \textbf{Speed A100} (ms)$\downarrow$ & $\theta$ (M)$\downarrow$ & \textbf{FLOP} (B)$\downarrow$ \\
        \specialrule{1.5pt}{1pt}{1pt}
        v8n & 640 & 37.3 & {\bf 80.4} & {\bf 0.99} & {\bf 3.2} & {\bf 8.7} \\
        v8s & 640 & 44.9 & 128.4 & 1.2 & 11.2 & 28.6 \\
        v8m & 640 & 50.2 & 234.7 & 1.83 & 25.9 & 78.9 \\
        v8l & 640 & 52.9 & 375.2 & 2.39 & 43.7 & 165.2 \\
        v8x & 640 & {\bf 53.9} & 479.1 & 3.53 & 68.2 & 257.8 \\
        \hline
    \end{tabularx}
    \label{yolo_family}
\end{table}


\section{Experiments and Results}
\label{sec:exp_res}

We conducted all of our experiments using Python and PyTorch~\cite{Torch_2016}. In this section, we demonstrate our experiments and results of testing models on unseen CT images.

\textbf{YOLO Model.}\label{ssec:YOLOv8n} It is based on the Ultralytics architecture~\cite{Jocher_2023} and specifically the nano version 8 model. It is acknowledged as the state-of-the-art for object detection, image classification, and segmentation tasks, incorporating numerous architectural enhancements compared to earlier models. To initialize the YOLOv8n model, we employed pre-trained weights trained on the COCO dataset. Subsequently, we fine-tuned the model using 100 kidney CT images. The training process consisted of a total of 30 epochs, with a fixed learning rate of 0.01. For optimization, we utilized the Adam algorithm with weight decay~\cite{Loshchilov_2018} (learning rate = 0.002, momentum = 0.9). The fine-tuning process on the NVIDIA Tesla T4 GPU was completed within 2 minutes. On average, it takes 286.5~ms to detect the ROI for the kidneys in each CT image.

\textbf{Image Processor.} The YOLOv8n model can detect both kidneys in most CT images in the test set. However, there are instances where only one kidney, three kidneys, or even no kidney is detected. Testing results are reported in Table~\ref{yolo_test_res}. It is important to note that no cases of more than three kidneys being detected were observed. The cases in which no kidney is detected are excluded from the kidney stone classification using CNN models. For the cases in which only one kidney is detected, we developed an Python image processor in an attempt to correct the prediction. This processor generates an additional traget region, represented by a red box, by horizontally mirroring the existing red bounding box. Only the first two predicted red boxes were used in cases where three kidneys were detected. Consequently, two kidneys were detected with red bounding boxes on all 323 CT test images. This resulted in the generation of 646 kidney images for the downstream CNN classification task.

\begin{table}[htbp]
    \caption{YOLOv8n model test evaluation on kidney detection task where 'KS' stands for Kidney Stone.}
    \vspace*{0.5em}
    \renewcommand{\arraystretch}{1.3}
    \begin{tabularx}{\linewidth}{
        >{\arraybackslash}p{0.42\linewidth}
        >{\centering\arraybackslash}p{0.08\linewidth}
        >{\centering\arraybackslash}p{0.08\linewidth}
        >{\centering\arraybackslash}p{0.08\linewidth}
        >{\centering\arraybackslash}p{0.08\linewidth}
        }
        \hline
        \centering\textbf{\# of detected kidneys $\rightarrow$} & \textbf{0} & \textbf{1} & \textbf{2} & \textbf{3} \\
        \specialrule{1.5pt}{1pt}{1pt}
        \# train images (with KS) & 0 & 0 & 61 & 2 \\
        \# train images (normal) & 0 & 0 & 127 & 10 \\
        \hline
        \# test images  (with KS) & 14 & 50 & 99 & 2 \\
        \# test images  (normal) & 9 & 41 & 131 & 0 \\
        \specialrule{1.5pt}{1pt}{1pt}
\centering\textbf{Total} & \textbf{23} & \textbf{91} & \textbf{418} & \textbf{14} \\
        \hline
    \end{tabularx}
    \label{yolo_test_res}
\end{table}

\textbf{MobileNet.}\label{ssec:MobileNet} In our study, the MobileNet classifier is a pre-trained model on ImageNet obtained from the PyTorch library~\cite{Torch_2016}. To align with our specific task of kidney stone detection, which involves binary classification, we replaced the original Softmax layer of the pre-trained MobileNet with a Sigmoid layer. For the training phase, we maintained a fixed epoch value of 50, a learning rate of $10^{-4}$, and a batch size of 4. Each epoch required approximately 30~s on a 12GB Intel(R) Haswell CPU. Both the training and test datasets employed a consistent batch size. The Adam optimizer was used during training, with a decaying learning rate of $10^{-4}$. Binary cross-entropy served as the loss function for binary classification. The training set consisted of 200 kidney images, which were augmented, resulting in a training set size of 293 kidney images. The test set consisted of 646 images. The average inference time of MobileNet running on the CPU is 0.62~s per image.

\subsection{Kidney Detection and Classification}
\label{ssec:detect_kidney}

\begin{figure}[htb]
  \includegraphics[width=\linewidth]{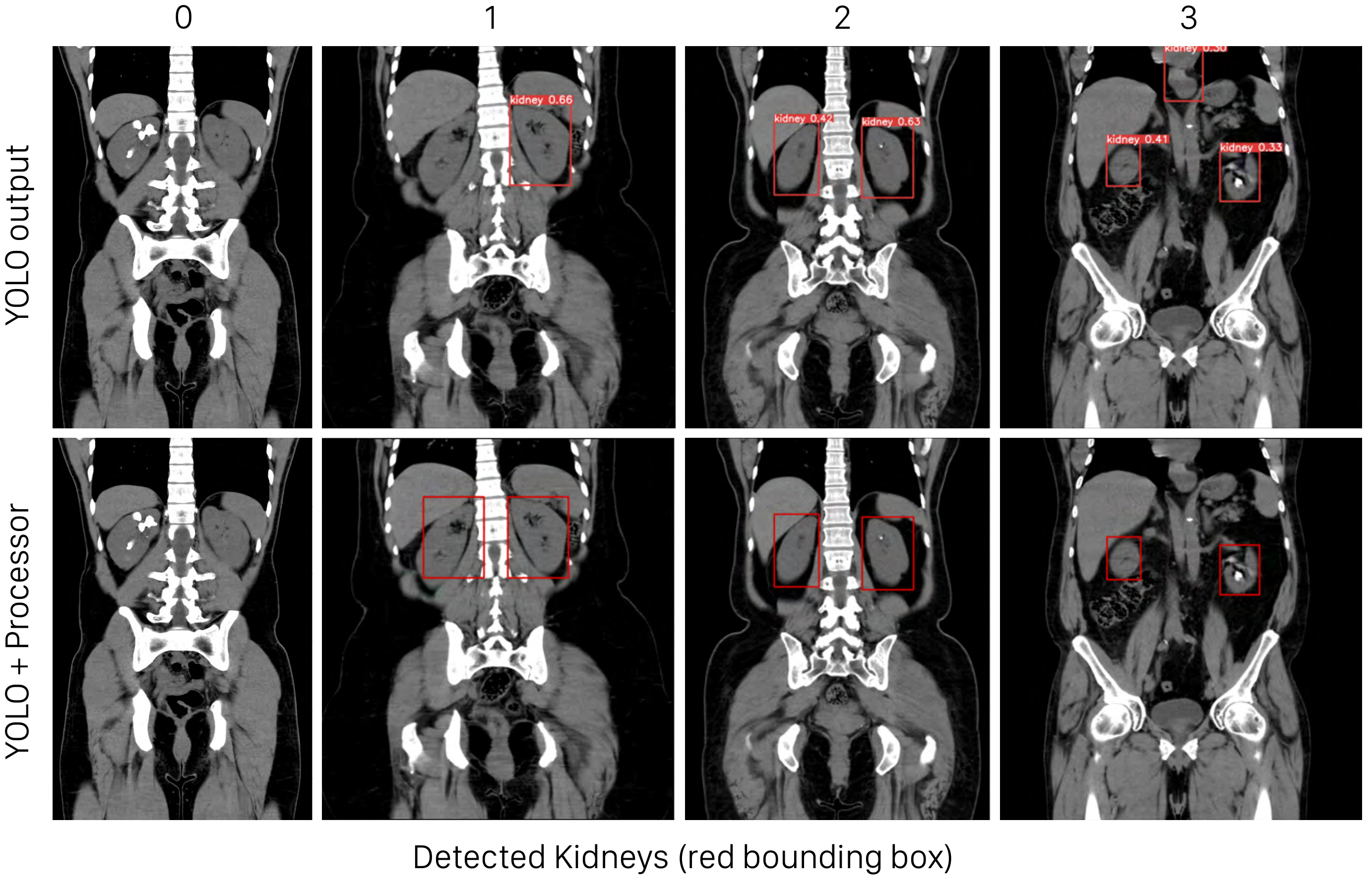}
    \caption{Kidney detection examples on test images using YOLOv8n only (1st row) and with prediction adjustment using our image processor (2nd row).}
\label{fig:kidney_pred}
\end{figure}

Fig.~\ref{fig:kidney_pred} illustrates examples of ROI detection in coronal CT images. Four distinct cases can occur during the detection process. However, since the detection of only two kidneys is necessary, our image processor can effectively manage these cases and produce the required ROI images as desired. This is illustrated in the second row of Fig.~\ref{fig:kidney_pred}.

It is noteworthy that data augmentation applied to stone-positive images does not yield significant benefits for kidney stone classification, as evidenced by the decrease in F1 score, Precision, and Recall. When utilizing the data augmentation technique, both the overall MobileNet model and its Fully Connected (FC) layer exhibit an increased False Negative (FN) rate, as indicated in Table~\ref{tab:cnn_test_res}. Additionally, fine-tuning the entire MobileNet yields better performance compared to fine-tuning only the FC layer. However, the gap in performance between the two approaches is small. Therefore, we argue that fine-tuning the FC layer, with optimized hyperparameters, can be a more efficient approach.

\begin{table}[ht]
    \caption{Performance of the classification model (MobileNetv2) evaluated on testing dataset. FC: MobileNet Fully Connected Layer. DA: Data Augumentation. FP: False Positive. FN: False Negative. P: Precision. R: Recall.}
    \vspace*{0.6em}
    \renewcommand{\arraystretch}{1.3}
    \begin{tabularx}{\linewidth}{
                >{\centering\arraybackslash}p{0.25\linewidth}|
                XXXXX
        }
        \hline
        & \textbf{F1 Score} & \textbf{P} & \textbf{R} & \textbf{FP Rate} & \textbf{FN Rate}\\
        \specialrule{1.5pt}{1pt}{1pt}
        ResNet-18 & 98\% & 98\% & 98\% & 2\% & 2\% \\
        \hline
        FC & 84\% & 84\% & 84\% & 16\% & 16\% \\
        FC+DA& 87\% & 87\% & 87\% & 13\% & 13\% \\
        \hline
        MobileNet & 98\% & 98\% & 98\% & 2\% & 2\% \\
        MobileNet+DA & \textbf{99\%} & \textbf{99\%} & \textbf{99\%} & \textbf{1\%} & \textbf{1\%} \\
        \hline
    \end{tabularx}
    \label{tab:cnn_test_res}
\end{table}


\section{Conclusions}
\label{sec:conclusions}

This study proposes a compact deep learning framework specifically designed for edge devices to assist doctors in the diagnosis of kidney stones. The framework comprises only 5.23 million parameters and has a size of 15~MB. Additionally, our proposed framework achieves 96\% accuracy in kidney stone detection using only 100 CT images. Furthermore, the results highlight that our framework achieves fast inference on a CPU, while still maintaining a competitive accuracy compared to state-of-the-art deep learning models. Our future work aims to investigate various modalities of data to assess the effectiveness of our framework, further optimize the framework for enhanced performance and achieve higher accuracy on edge devices.

\section{Acknowledgments}
\label{sec:acknowledgments}

This publication has emanated from research conducted with the financial support of Science Foundation Ireland under Grant number 18/CRT/6183. For the purpose of Open Access, the author has applied a CC BY public copyright licence to any Author Accepted Manuscript version arising from this submission.

\bibliographystyle{IEEEbib}
\bibliography{strings}

\begin{thebibliography}{10}

\bibitem{Penniston_2012}
Kristina~L Penniston and Stephen~Y Nakada,
\newblock ``Development of an instrument to assess the health related quality of life of kidney stone formers,''
\newblock {\em J Urol}, vol. 189, no. 3, pp. 921--930, Sept. 2012.

\bibitem{Chewcharat_2021}
Api Chewcharat and Gary Curhan,
\newblock ``Trends in the prevalence of kidney stones in the united states from 2007 to 2016,''
\newblock {\em Urolithiasis}, vol. 49, no. 1, pp. 27--39, Feb. 2021.

\bibitem{Turk_2015}
Christian T{\"u}rk, Ale{\v s} Pet{\v r}{\'\i}k, Kemal Sarica, Christian Seitz, Andreas Skolarikos, Michael Straub, and Thomas Knoll,
\newblock ``{EAU} guidelines on diagnosis and conservative management of urolithiasis,''
\newblock {\em Eur Urol}, vol. 69, no. 3, pp. 468--474, Aug. 2015.

\bibitem{Shen_2017}
Dinggang Shen, Guorong Wu, and Heung-Il Suk,
\newblock ``Deep learning in medical image analysis,''
\newblock {\em Annu. Rev. Biomed. Eng.}, vol. 19, no. 1, pp. 221--248, June 2017.

\bibitem{Hesamian_2019}
Mohammad~Hesam Hesamian, Wenjing Jia, Xiangjian He, and Paul Kennedy,
\newblock ``Deep learning techniques for medical image segmentation: Achievements and challenges,''
\newblock {\em Journal of Digital Imaging}, vol. 32, no. 4, pp. 582--596, Aug. 2019.

\bibitem{talo_2019}
Muhammed Talo, Ulas~Baran Baloglu, Özal Yıldırım, and U~Rajendra~Acharya,
\newblock ``Application of deep transfer learning for automated brain abnormality classification using {MR} images,''
\newblock {\em Cognitive Systems Research}, vol. 54, pp. 176--188, May 2019.

\bibitem{ozturk_2020}
Tulin Ozturk, Muhammed Talo, Eylul~Azra Yildirim, Ulas~Baran Baloglu, Ozal Yildirim, and U.~Rajendra~Acharya,
\newblock ``Automated detection of {COVID}-19 cases using deep neural networks with {X}-ray images,''
\newblock {\em Computers in Biology and Medicine}, vol. 121, pp. 103792, June 2020.

\bibitem{Kijowski_2020}
Richard Kijowski, Fang Liu, Francesco Caliva, and Valentina Pedoia,
\newblock ``Deep learning for lesion detection, progression, and prediction of musculoskeletal disease,''
\newblock {\em Journal of Magnetic Resonance Imaging}, vol. 52, no. 6, pp. 1607--1619, Dec. 2020.

\bibitem{hirschorn_2014}
David~S. Hirschorn, Asim~F. Choudhri, George Shih, and Woojin Kim,
\newblock ``Use of {Mobile} {Devices} for {Medical} {Imaging},''
\newblock {\em Special Bonus Issue for 2014: ACR Imaging IT Reference Guide}, vol. 11, no. 12, Part B, pp. 1277--1285, Dec. 2014.

\bibitem{Redmon_yolo_2016}
J.~Redmon, S.~Divvala, R.~Girshick, and A.~Farhadi,
\newblock ``You only look once: Unified, real-time object detection,''
\newblock in {\em 2016 IEEE Conference on Computer Vision and Pattern Recognition (CVPR)}, Los Alamitos, CA, USA, jun 2016, pp. 779--788, IEEE Computer Society.

\bibitem{Jocher_2023}
Glenn Jocher, Ayush Chaurasia, and Jing Qiu,
\newblock ``Yolo by ultralytics,'' Jan. 2023.

\bibitem{Yildirim_2021}
Kadir Yildirim, Pinar~Gundogan Bozdag, Muhammed Talo, Ozal Yildirim, Murat Karabatak, and U~Rajendra Acharya,
\newblock ``Deep learning model for automated kidney stone detection using coronal {CT} images,''
\newblock {\em Comput Biol Med}, vol. 135, pp. 104569, June 2021.

\bibitem{Jou_2016}
Brendan Jou and Shih-Fu Chang,
\newblock ``Deep cross residual learning for multitask visual recognition,''
\newblock in {\em Proceedings of the 24th ACM International Conference on Multimedia}, New York, NY, USA, 2016, MM '16, p. 998–1007, Association for Computing Machinery.

\bibitem{Islam_2022}
Md~Nazmul Islam, Mehedi Hasan, Md~Kabir Hossain, Md~Golam~Rabiul Alam, Md~Zia Uddin, and Ahmet Soylu,
\newblock ``Vision transformer and explainable transfer learning models for auto detection of kidney cyst, stone and tumor from {CT-radiography},''
\newblock {\em Scientific Reports}, vol. 12, no. 1, pp. 11440, July 2022.

\bibitem{liu_2021}
Ze~Liu, Yutong Lin, Yue Cao, Han Hu, Yixuan Wei, Zheng Zhang, Stephen Lin, and Baining Guo,
\newblock ``Swin transformer: Hierarchical vision transformer using shifted windows,''
\newblock in {\em Proceedings of the IEEE/CVF International Conference on Computer Vision (ICCV)}, 2021.

\bibitem{Simonyan_2015}
Karen Simonyan and Andrew Zisserman,
\newblock ``Very deep convolutional networks for large-scale image recognition,''
\newblock in {\em International Conference on Learning Representations}, 2015.

\bibitem{he_2016}
Kaiming He, Xiangyu Zhang, Shaoqing Ren, and Jian Sun,
\newblock ``Deep residual learning for image recognition,'' 2016.

\bibitem{Szegedy_2015}
C.~Szegedy, Wei Liu, Yangqing Jia, P.~Sermanet, S.~Reed, D.~Anguelov, D.~Erhan, V.~Vanhoucke, and A.~Rabinovich,
\newblock ``Going deeper with convolutions,''
\newblock in {\em 2015 IEEE Conference on Computer Vision and Pattern Recognition (CVPR)}, Los Alamitos, CA, USA, jun 2015, pp. 1--9, IEEE Computer Society.

\bibitem{baygin_2022}
Mehmet Baygin, Orhan Yaman, Prabal~Datta Barua, Sengul Dogan, Turker Tuncer, and U.~Rajendra Acharya,
\newblock ``Exemplar {Darknet19} feature generation technique for automated kidney stone detection with coronal {CT} images,''
\newblock {\em Artificial Intelligence in Medicine}, vol. 127, pp. 102274, May 2022.

\bibitem{Bhandari_2023}
Mohan Bhandari, Pratheepan Yogarajah, Muthu~Subash Kavitha, and Joan Condell,
\newblock ``Exploring the capabilities of a lightweight cnn model in accurately identifying renal abnormalities: Cysts, stones, and tumors, using lime and shap,''
\newblock {\em Applied Sciences}, vol. 13, no. 5, 2023.

\bibitem{Metser_2009}
Ur~Metser, Sangeet Ghai, Yang~Yi Ong, Gina Lockwood, and Sidney~B Radomski,
\newblock ``Assessment of urinary tract calculi with {64-MDCT}: The axial versus coronal plane,''
\newblock {\em American Journal of Roentgenology}, vol. 192, no. 6, pp. 1509--1513, June 2009.

\bibitem{Sandler_2018}
Mark Sandler, Andrew~G. Howard, Menglong Zhu, Andrey Zhmoginov, and Liang-Chieh Chen,
\newblock ``Mobilenetv2: Inverted residuals and linear bottlenecks,''
\newblock {\em 2018 IEEE/CVF Conference on Computer Vision and Pattern Recognition}, pp. 4510--4520, 2018.

\bibitem{Olga_2015}
Olga Russakovsky, Jia Deng, Hao Su, Jonathan Krause, Sanjeev Satheesh, Sean Ma, Zhiheng Huang, Andrej Karpathy, Aditya Khosla, Michael Bernstein, Alexander~C. Berg, and Li~Fei-Fei,
\newblock ``{ImageNet Large Scale Visual Recognition Challenge},''
\newblock {\em International Journal of Computer Vision (IJCV)}, vol. 115, no. 3, pp. 211--252, 2015.

\bibitem{lin_2014}
Tsung-Yi Lin, Michael Maire, Serge Belongie, Lubomir Bourdev, Ross Girshick, James Hays, Pietro Perona, Deva Ramanan, C.~Lawrence Zitnick, and Piotr Dollár,
\newblock ``Microsoft coco: Common objects in context,'' 2014,
\newblock cite arxiv:1405.0312Comment: 1) updated annotation pipeline description and figures; 2) added new section describing datasets splits; 3) updated author list.

\bibitem{Mahmud_2023}
Sakib Mahmud, Tariq~O. Abbas, Adam Mushtak, Johayra Prithula, and Muhammad E.~H. Chowdhury,
\newblock ``Kidney cancer diagnosis and surgery selection by machine learning from ct scans combined with clinical metadata,''
\newblock {\em Cancers}, vol. 15, no. 12, 2023.

\bibitem{Torch_2016}
TorchVision maintainers and contributors,
\newblock ``Torchvision: Pytorch's computer vision library,'' Nov. 2016.

\bibitem{Loshchilov_2018}
Ilya Loshchilov and Frank Hutter,
\newblock ``Decoupled weight decay regularization,''
\newblock in {\em International Conference on Learning Representations}, 2019.

\end{thebibliography}

\end{document}